# A new method of brain stimulation at ultra-high frequency

Yousef Jamali[1*], Mohammad Jamali[2,3], Mehdi Golshani[2,3]

1) Department of Applied Mathematics, School of Mathematical Sciences, Tarbiat Modares University, Tehran, Iran
2) Department of Physics, Sharif University of Technology, Tehran, Iran
3) School of Physics, Institute for Research in Fundamental Sciences (IPM), Tehran, Iran

* Correspondence: y.jamali@ipm.ir

## Summary

Nerve stimulation via micro-electrode implants is one of the neurostimulation approaches which is used frequently in the medical treatment of some brain disorders, neural prosthetics, brain-machine interfaces and also in the cyborg. In this method, the electrical stimulation signal can be categorized by the frequency band: low frequency, high frequency, and ultra-high frequency. The stimulation should be less destructive, more smooth, and controllable. In this article, we present a brief description of the mechanism underlying the ultra-high frequency stimulation. In the flowing, from an informatics perspective, we propose a state-of-the-art, low destructive, and highly efficient stimulation method at the low amplitude ultra-high frequency signal. In this method, we have tried to reduce the adaptation of the nerve system by modulating the stimulation signal via a low frequency rectangular random wave. By this method, we could reach the "almost zero discharge" with minimum destructive effect in the experimental test on the fish nervous system.

# Introduction

The purpose of electrical nerve stimulation (neurostimulation), that includes implanting electrodes in the nerve system and applying an electric field to it, is pushing neurons to fire. Many factors, such as the shape and material of electrodes [1-11], implanted area (including its symmetry and asymmetry, etc. [12-14]), the property of signal (such as: shape, amplitude, and frequency) [1, 15-21] and the type of stimulation (i.e. constant voltage, constant current or constant charge [1, 5, 15, 22, 23]) can be affected the quality of stimulation.

The adverse effects of direct current stimulation (DCS) or long-pulsed current stimulation (PCS), such as tissue and electrode damaging [1, 3], making them inappropriate for nervous system stimulates. Hence caution should be exercised when using these methods in some sensitive applications, such as the deep brain stimulation (DBS). Of course, this method is still useful for the blocking of the nerve system [24, 25].

In the neurostimulation, applied frequency bands (range, spectrum) could be categorized into a lower frequency (<100Hz) [8, 24-34] higher frequency (100-1000 Hz) [4, 7, 9, 11, 35] and ultra-high frequency (>500 Hz) [36-38]. Low-frequency (The lower band) stimulation is used in most of the works since the physiological frequency of neurons typically is less than 100 Hz [37] (though it could reach up to 700 Hz in some cases [36, 39]). This physiological restriction of frequency is related to three facts; first: the existence of a refractory period following each action potential which is in the order of milliseconds; second: limitation in the speed of synaptic action, and third; temporary depletion of synaptic vesicles at a persistent high (>100Hz) frequency stimulation (synaptic fatigue) [36, 37, 40-43].

Over recent years, the use of high-frequency bipolar stimulation (HFBS) has been extensively increased. This tendency is due to some factors such as: reduction of adverse physiological and electrochemical effects at HFBS [1, 3, 4, 16, 44-47], the experimental confirmation of the possibility of nerve stimulation at high frequency (~kHz) [37] (perhaps, because of the accumulation effect [37, 48, 49]) and variety of effective parameters at the HFBS [37]. Although the experimental results indicate the successfulness of this method [37, 50] there is no global, one-size-fits-all model to explain the mechanism of stimulation in this frequency range[36, 37]. Different models have been used in order to explain some aspects of the behaviors.

Ultra high frequency has been mainly used for three cases: nerve conduction blocking, radiofrequency nerve ablation (and heating), and nerve stimulation. In the nerve conduction blocking, which used for example in chronic pain treatment [36, 37, 51-54], stimulation signal in the kHz-frequency (1-100 KHz) is continuously applied to the tissue. This stimulation prevents signal transmission through the nerve, by blocking axons conduction [36].

The radiofrequency (>500KHz) nerve ablation, by heating up a small area of nerve tissue, eliminates the nerve and cancer tissue or relieves the pain (because of decrease in the nervous system's activity at high temperatures)[3]. In the nerve stimulation, which is mainly used in electrotherapy [37] and cyborg [7, 55], the ultra-high frequency is modulated by a low-frequency carrier signal (50-100 Hz). A variety of modulation techniques, such as interference of multi different high-frequencies or using carrier square pulse, have been developed [37].

In the last application (nerve stimulation), after applying an ultra-high frequency stimulation, a brief onset response, followed by an exponential decay of the firing rate, have been observed experimentally and studied theoretically [36, 56].

To prevent this decay, the signal could be switched off and on periodically. The off state gives enough time for the nerve cells to recover and prepare them for the next stimulation. Hence the switching frequency's choice is based on the recovery time and is in the order of natural frequency of firing (Figure 1).

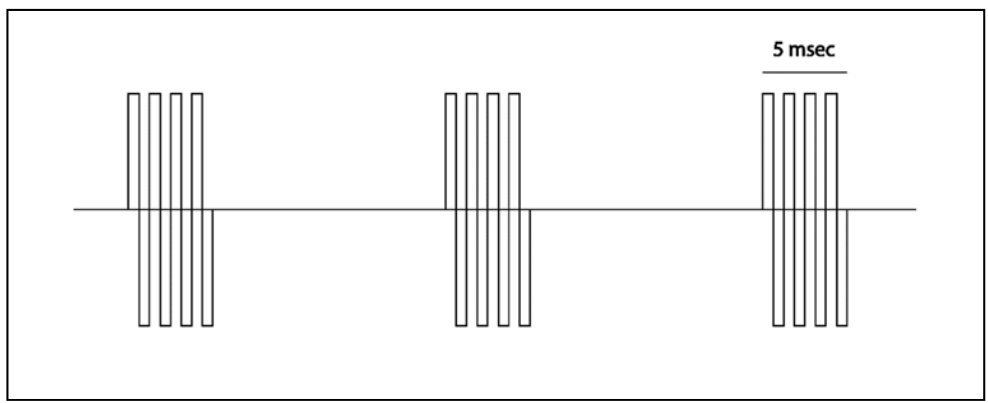

*Figure 1. a stimulation signal at an ultra-high frequency that switches on & off at low frequency.*

In this article, after the investigation of high-frequency stimulation mechanism, our aim is to find the optimal stimulation (which is more effective with less signal strength). We shall show that neuronal ultra-high frequency stimulation, when modulated on with a random time-varying low-frequency carrier, will result in an optimal stimulation with lower adverse effects. To confirm our hypothesis, some fish brains were stimulated electrically in vivo, using the aforementioned method and we compared it with other methods in terms of its charge injection (charge and discharge amplitude) per cycle and fish locomotion respond.

## Ultra-high frequency electrical stimulation

As we outlined above, stimulation at an ultra-high frequency (3-500KHz) and with an intensity which would not trigger the nerve blocking [36], leads to the initiation of neuronal firing. However, the firing rate decays exponentially to zero with a time constant of multi-seconds [39, 57-61]. Hence applying this stimulation constantly, if doesn't cause blocking ( nerve conduction block or neurotransmitter depletion block [36, 37, 40-43]) causes an onset respond proportional to the intensity of the excitation current. However, this response is transient and will not last, as shown schematically in Figure 2.

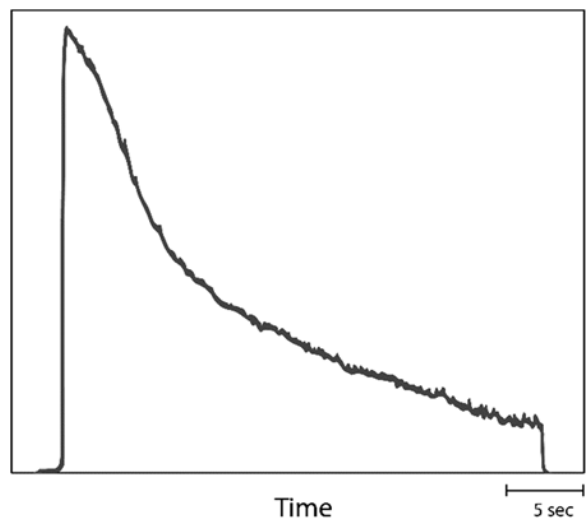

*Figure ٢. schematic pattern of the onset response of a neuron to ultra-high frequency stimulation with low intensity. The picture is adapted from [36]*

After stopping the stimulation (i.e. turning off the neurostimulator), the organism starts to recover [The recovery time depends on the stimulation parameters, such as frequency, amplitude, and shape, as will be discussed in details in the following.] and returns to its normal mode. Therefore, by alternately connecting and disconnecting an appropriate very high-frequency stimulation by means of enough low-frequency signal, it is possible to achieve a long-lasting and steady nerve system response. There are some common methods which use this approach:

1. Using a superposition current [37, 62-68]. This method, as is shown in Figure 3a, involves the application of two currents (usually ~ 2.5 KHz sinusoidal signal) at slightly different frequencies (~50 Hz). These two waves interfere with each other and create a modulated low-frequency signal of ~50 Hz. The wavelets will repeat at regular intervals.

2. Using a modular square current [37, 66-70]. This involves switching on and off of a high-frequency square/sinusoidal signal (1-100KHz). The switching is driven by a square low-frequency signal (50Hz), with an adjustable duty [37] (Figure 3b). This type of stimulation, due to the duty control, is less destructive during a short stimulation and is more efficient.

3. Using a high-frequency stimulation with non-zero offset. In the ref. [71] a burst sinusoidal high frequency, with a non-zero offset, was used for stimulation (Figure 3c). In addition, in the ref. [72], authors suggested a voltage control stimulus which is switched on and off repeatedly, with a high frequency (upto100kHz) duty cycled signal (non-zero offset).

4. Using a low-frequency rectangular waveform stimulation with low Duty-cycle (Figure 3d). This type of stimulation is classically placed into the current pulse (PC) stimulation category. The fact of the existence of high-frequency components in its Fourier spectrum makes it belong to the high-frequency stimulation in order of $1/T$ which $T$ is the pulse duration. Especially, in the constant voltage stimulation, due to the capacitive property of environment and electrodes, the shape of current is a bipolar periodic wave with the same pulse duration (Figure 4). Recently this method has been widely used in the DBS [16, 47, 73-79]. Because of low Duty-cycle, it, too, would have little adverse electrochemical effects [1, 3, 47].

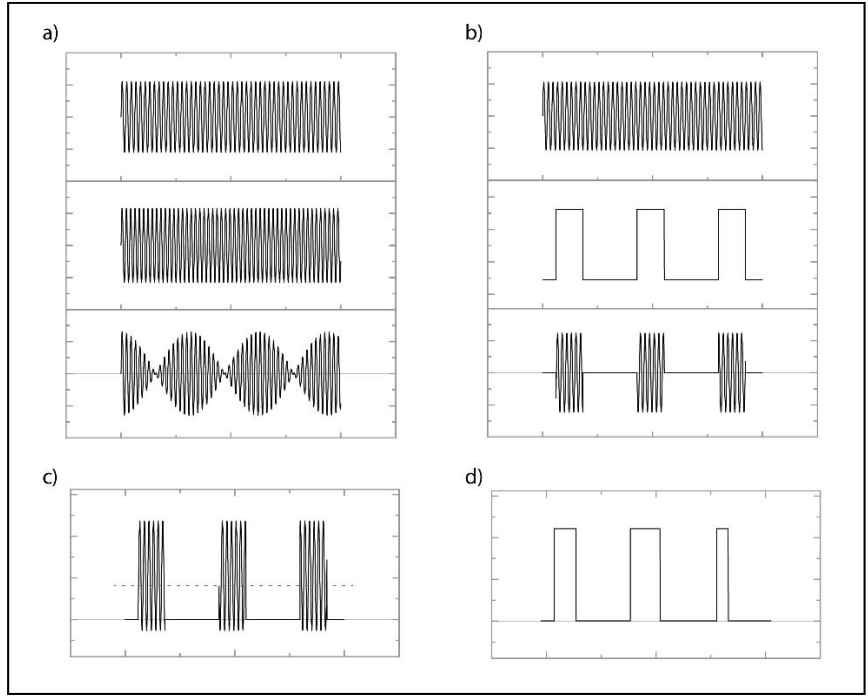

*Figure ٣ : By alternately connecting and disconnecting an appropriate very high-frequency stimulation by means of enough low-frequency signal, it is possible to achieve a long-lasting and steady nerve system response. There are some common methods which use this approach: a) Using a superposition current; b) Using a modular square current; c) Using a high-frequency stimulation with non-zero offset; d) Using a low-frequency rectangular waveform stimulation with low Duty-cycle.*

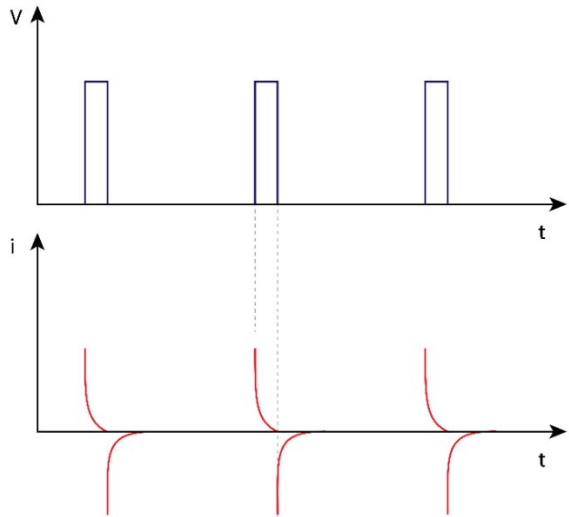

*Figure 4: In the constant voltage stimulation (top figure), due to the capacitive property of environment and electrodes, the shape of current is a bipolar periodic wave with zero charge balance (down figure).*

Using bipolar alternative stimulation, with zero charge balance, could offer stimulation with less adverse effects. Table 1 shows some examples of applied stimulation, with the zero charge balance.

## Biophysical mechanism of high-frequency stimulation effects

The reaction of the nerve system to the stimulation methods outlined above consists of two stages. The first stage is an initial respond of neurons to the ultra-high frequency stimulation, and the second one is the recovery and returns to the resting state. Both stages are discussed in more detail below.

### The first stage:

As the transmembrane field is increased, due to the right external electric field, the opening probability of a given voltage-gated sodium channel is raised. Consequently, the number of open channels is increased. Due to the existence of an asymmetry in the closing and opening dynamics of channels, for an open channel, by removing the external field or even applying the inverse one, these channels will not be closed (by the same opening mechanism). Hence if the field amplitude and the frequency are high enough [51, 52] so that the summation phenomena occur [37, 48, 49], after certain periods, the induced membrane fields reach the threshold and initiate an action potential, (Figure 5). This onset response, both in-vitro [37, 48, 51, 52] and in-silico [53, 54, 80, 81] have been well characterized. It's worth noting that according to the nerve membrane time constant ($\tau = RC \sim 10\ ms$), the stimulation frequency must be more than $\frac{1}{\tau}$, which is in the high frequency range.

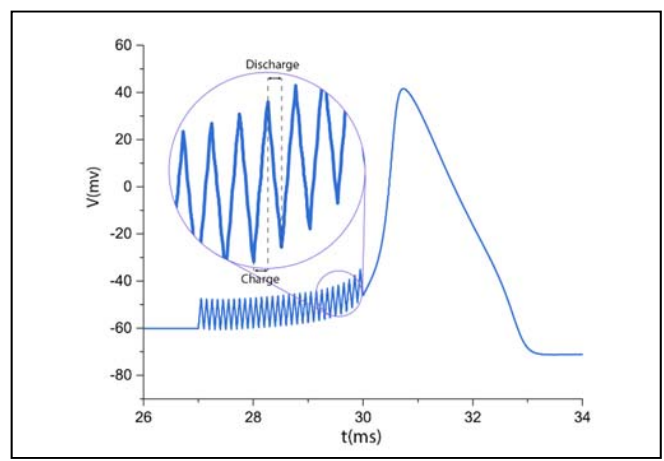

*Figure 5. Simulation result of neuron responds and variation of membrane potential under high-frequency stimulation based on the H&H dynamics model [our unpublished simulation result].*

In addition to the axon's characteristics, like diameter and length, the intensity of neuron's firing depends on the frequency and the amplitude of applied field [36, 39]. In principle, increasing the frequency of stimulation increases stimulation or, in other words, results in more efficient stimulation (efficient stimulation means reducing the threshold of the applied field intensity of stimulation neurons or more stimulation in a determined intensity). The main reasons for this is, on the one hand, the asymmetry in the dynamics of channel activation based on the H&H equations (Figure 5 shows the simulation result of neuron dynamics under H&H dynamics), and on the other hand, an increase in the effective field (the transmembrane fields resulting from the applied filed) which causes the rise of opening probability of sodium channels [our unpublished simulation result]. This will be discussed in the next session.

### The effect of an induced electric filed on raising the probability of firing

In the high-frequency stimulation, if the field amplitude is decreased (due to the reduction in the opening probability of sodium channels), while the closing rate is constant and is independent of field amplitude, then sodium current decreases. Hence, it can result in a lack of the summation process [1]. In other words, the effective time and the amplitude of stimulation should be large enough, so that the opening rate of sodium ion channels overcome their closing rate. In this situation, the system leaves its resting state and goes to the firing mode. Nevertheless, in the low amplitude situation, although the field itself does not directly cause neurons to fire, it does change and even elevates the transmembrane potential. This elevation, in turn, results in increasing the neuron sensitivity to other external stimulations. Thus, firing likelihood goes up, compared to the normal situation.

As we know, neurons have a baseline firing rate in-vivo, due to different parameters, such as random synapse current and/or environmental noise. Hence the above mentioned increased sensitivity results in the elevation of baseline firing rate. In other words, even at low amplitudes, for which no direct stimulation can occur, the baseline rate of neurons will increase, because of their higher sensitivity.

Hence, applying a low amplitude field could be equal to an increase in the response of neurons to the environmental stimulation, especially the environmental noise. As a result, the baseline firing rate of a neuron under such stimulation is higher compared to other neurons. In the following section, we shall outline how increasing the duration of an applied field, regardless of the summation property, could increase the probability of stimulation, especially at symmetric high frequencies.

### Increasing the effective time via ultra-high frequency stimulation.

Just after applying electrical stimulation, ions around the electrodes move and a plane of charge, at the surface of the metal electrode, is opposed by a plane of opposite charge, as "counterions", in the electrolyte. This double layer capacitor around the electrodes [1, 3, 82-84], as well as shielding of electrodes by polar molecules, such as water and polar proteins, results in induced electric fields decaying exponentially to zero with a time constant of 10 microseconds [1, 3]. Hence neurons around the electrode sense the stimulation field only in the first few microseconds, which we term as effective time [In fact, what we explain here is just a non-faradaic current. There is two types of currents, Capacitive current, which is also called "non-faradaic" or "double-layer" current, does not involve any chemical reactions. It only causes accumulation (or removal) of electrical charges on (or from) the electrode and in the electrolyte solution near the electrode. This type of current is reversible and nondestructive. On the other hand, the faradaic current is the current generated by the reduction or oxidation of some chemical substance at an electrode. The net faradaic current is the algebraic sum of all the faradaic currents flowing through an electrode. This type of current is destructive and we prefer reducing it as much as possible. At low amplitude electric fields, the faradaic current is negligible]. When the polarity of the applied voltage source is reversed, the direction of current is reversed, the charge redistribution is reversed, and the charge that was injected from the electrode into the electrolyte and stored by the capacitor may be recovered. As a result, by increasing the frequency of the charge balanced stimulation, it is possible to reach a longer total effective time in the defined period of time.

In both capacitive [1-3] and voltage-controlled (potentiostatic) stimulation methods [1], which are in the electrochemical reversibility range [85, 86] [Several experimentalists have made the choice of using voltage-controlled stimulations to remain within the reversible range of the electrodes (Joucla and Yvert, 2012).], the implementation of a square wave stimulation provides a decreasing exponential current in the tissue and electrode (the faradaic current can be removed by adding an external serial capacitor to the circuit). Hence in accordance with quasi-static or quasi-stationary equations [18, 19, 87-90], an intense current, at the initial moments of stimulation ($\sim 10\ \mu s$), results in an induced electric field equal to:

$$|E| = \frac{I_0 \omega}{2\pi\sqrt{\sigma^2 + \epsilon^2\omega^2} R} + \frac{I_0 l \omega}{2\pi\sqrt{\sigma^2 + \epsilon^2\omega^2} R^3}$$

$$|E| = I_0 \omega \left(\frac{\alpha}{R} + \frac{\beta}{R^3}\right)$$

This strong electric field belongs to the transient state of the electrode-tissue circuit. However, by increasing the stimulation frequency and repeating this situation, it is possible to stimulate the nerve system and changing the state of ion channels.

It is clear that a larger induction field intensity causes a stronger stimulation, but the electrochemical detrimental effects of current controls the upper bound of intensity. Our data and other experimental data [1, 73-79, 90] show the positive effects of increasing frequency for the reduction of the injected charge amplitude, achieving a proper stimulation of neurons. In addition, experimental evidence [37, 51, 52] show that the type of the axon (neuron) could affect the stimulation threshold, as will be discussed in more detail in the following section.

## Onset response and stimulation intensity

Experimental data indicate that the high-frequency stimulation, by both in blocking and stimulating neurons [36, 39, 48, 49, 57-61, 80], results in the "onset response", which is the transitory volley of activity produced in the nerve (Figure 2). There are two suggestions to reduce this undesirable phenomenon. First, via smoothing the stimulation current [36] which causes intensification of the destructive faradaic current [91]. The second one is the reduction of the stimulation intensity, e.g. by applying a lower amplitude field [36]. The drawback of decreasing the applied field amplitude is the reduction in the nerve response to stimulation. However, as mentioned above, by using a more square waveform shape stimulation and increasing its frequency, it is possible to overcome this problem and to achieve a weak but effective stimulation.

## Second Stage:

In continuous ultra-high frequency stimulation, many neurons show a reduction in the firing frequency of their spike response, following an initial increase (Figure 2). This happens because of several phenomena, such as spike-frequency adaptation [36], fast neural adaptation, and synaptic fatigue [37, 40-43].

This undesirable behavior can be reduced by stopping the stimulation and giving neurons enough time (recovery time) to return to their normal condition and then repeating the stimulation.

Recovery time depends on the inherent properties of stimulated neurons (equivalent RC time) and stimulation parameters, such as amplitude. This process, i.e. switching stimulation off and on periodically in the natural frequency of neurons, is the basic principle of the stimulation at ultra-high frequencies.

Figure 6 which was adapted from the references [51, 52], shows the dependency of recovery time to stimulation amplitudes. It is worthy to note that in the low amplitude stimulation, the recovery time is more a function of the size and type of excitable neurons [56].

# The effect of the shape and type of axon in high-frequency stimulation

The relation between axon diameter and the size of the effective surface of excitation, and the intrinsic electrical properties of the axon, highlights the role of axonal diameter in the ultra-high frequency stimulation in two ways:

1. The experimental results [51, 52] show that the recovery time is directly related to the minimum number of cycles in each stimulation period. Hence based on the length of recovery time, the length of the stimulation period is specified. This suggests that the recovery time determines the switching frequency.

2. It is well known that there is an inverse correlation between refractory period [this is the time during which another stimulus given to the neuron (no matter how strong) would not lead to a second action potential.] and axon's diameter. In fact, the distal absolute refractory period was inversely correlated with distal conduction velocity. As a result, the axon diameter is one of the parameters which controls the upper limit of the firing rate [56]. Thus, larger diameter could cover a larger range of firing frequency. Hence the larger the axon's diameter, the shorter the duration of the action potential and refractory period, and hence the larger the maximal frequency of firing [92, 93]. The authors suggest a relationship between the axon diameter and the information rate so the axon diameter would be the main determinant of the properties of signal propagation along such axons.

Hence, increasing the neuron diameter, along with rising the time of recovery and neuron rest, results in neuron firing at the lower cycle number of stimulation at a constant intensity. It also causes an increase in the frequency range capacity of the random frequency stimulation. Results from both the empirical data [37, 51, 52] and the simulation [our unpublished simulation result], support the idea of stimulation in the lower cycles, by increasing of the neuron diameters in the pain, motor, and sensory neurons.

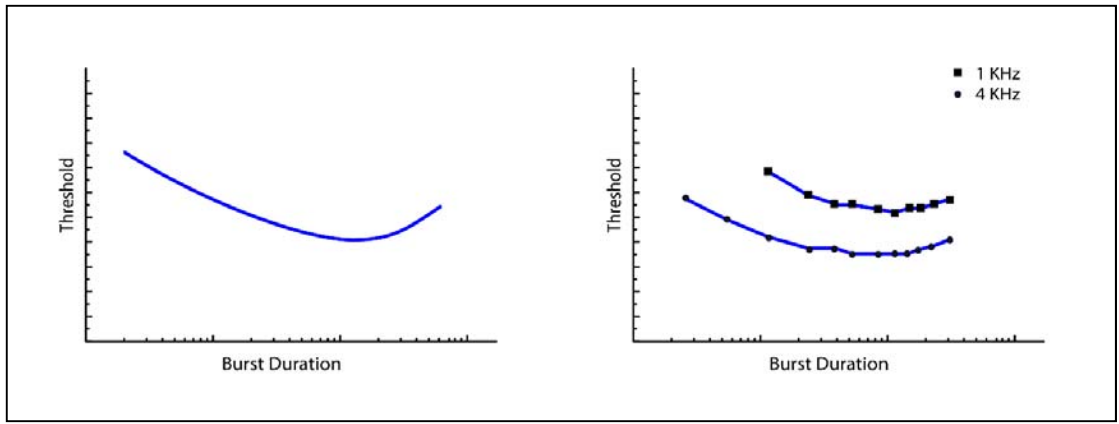

*Figure 6. Diagrams indicate the relation between the threshold and minimum burst duration of stimulation that result in neuron firing. this figure is a schematic of the result of the references [51, 52].*

this could be extended as a rule that: for thinner neurons, the stimulation threshold is increased [51]. In addition, due to different recovery time in the motor, sensory and pain neurons, the switching frequency is different and is a function of the neuron type [51]. Hence it seems that by tuning switching frequency, it is possible to optimize stimulation for a specific neuron type. However, our results still show a reduction in the central nervous system's response to the ultra-high frequency stimulation, for which it is proportional to the frequency. This new type of reduction could be termed behavioral reduction or informatics reduction and is discussed in more detail in the following.

# Neural informational adaptation

Neural adaptation is a change over time in the responsiveness of the nerve system to a constant stimulus. By this phenomenon, even with an ultra-high frequency, modulated by a low-frequency carrier signal, the nerve system response decays gradually. Interestingly, our result shows that, like fatigue effect [37, 40-43], the rate of decay is proportional to the frequency in the central nerve system. Delivered information integrates and processes in the CNS. According to the authors' speculation, how CNS process analyzes and evaluates this information and what information is transmitted during stimulation could be inspiring in designing new experiments and new methods for getting more efficient stimulation. In principle, the symmetrical structure of the stimulus signal (which have a low entropy, and contain little information) causes a uniform and symmetrical stimulation of neurons. This uniformity of stimulation markedly reduces the time of CNS adaptation [As G. Ellis et all have mentioned in page 308 of his book titled 'How can physics underlie the mind?'[94]: "Animals learn to distinguish events that occur regularly together from those that are only randomly associated. Thus, the brain has evolved a mechanism that 'makes sense' out of events in the environment by assigning a predictive function to some events, based on past experience and learned rules of behavior [95]. This predictive capacity is built into the continually adapting connections in neural networks in the brain [96]. Each region of the cortex has a repertoire of sequences it knows, and it has a name for each sequence it knows[96] p. 129. And the known must be distinguished from what is new…"]. Hence, by variation of the stimulus signal, it is possible to increase the input information and reduce the adaptation rate. On the other word, for the same amplitude, this method could get a more efficient stimulus in respect of neural total respond and adaptation (especially in the sensory neurons which have wider effective frequency capacity [56]). Given that, random (informational) packets can carry more information [97]. Our suggestion for more effective stimulation is to create a high-frequency stimulation, modulated with a square-wave modulator, whose frequency is random, with Poisson distribution and neuron physiological frequency mean. (Figure 7)

Our result shows that at a high-frequency stimulus, which is switched on and off randomly according to a Poisson distribution, with mean 100Hz and 10% duty, even a very low discharge in each period ($\sim 0.3\ nc$) results in a nerve stimulation and a macroscopic response of the subject. It is worth to note that with the same stimulation conditions, but with a uniform switching frequency, we didn't see any macroscopic response in the subject.

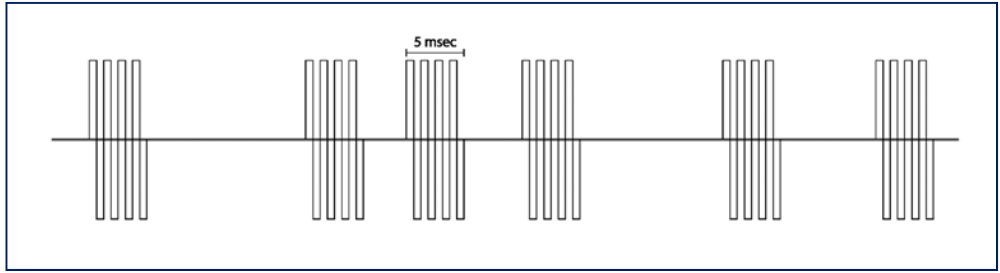

*Figure 7. A high-frequency stimulation, modulated with a square-wave modulator whose frequency is random with a Poisson distribution and a neuron physiological frequency mean.*

In addition, due to the variety of axons and their frequency carrier capacity [56], we speculate that the stimulation with random high frequency, with regular/random switching on and off, can utilize the neuron frequency capacity more efficiently and is also suitable for thicker neurons, such as sensory or brain neurons due to their broad firing frequency coverage. We didn't perform any experiment around this idea, but in this context, may be the result of ref [98] could be justified. The rectangular signal stimulation of this research includes the random pulse in duration or amplitude. Since the Fourier transform of the rectangular pulse has a wide range of frequency proportional to the inverse of duration and considering its randomness, this method would be justified in the above-mentioned context (the random high frequency with regular switching) (Figure 8). Also, random stimulation amplitude results in random firing rate response of neurons.

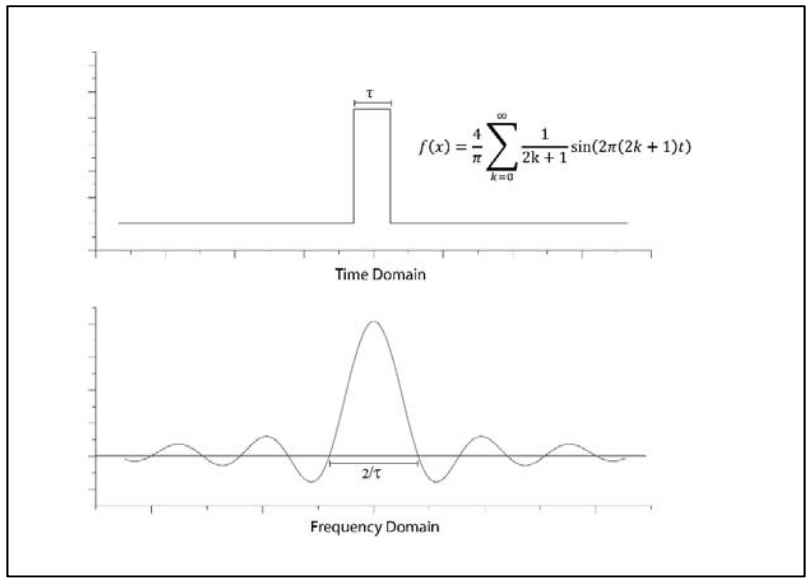

*Figure 8. Fourier transform of a rectangular pulse. As the pulse duration is made shorter the signal frequency range becomes wider and it includes the higher frequency component.*

# Our data on random stimulation

All experimental procedures were in accordance with the guidelines of the National Institutes of Health and the Iranian Society for Physiology and Pharmacology and were approved by the animal care and use committee of the Institute for Research in Fundamental Sciences.

1. We had the stimulation of clarias gariepinus motor neuron (control system) located in front of the cerebellum. As we reported in detail in [99], for testing the above hypothesis we stimulated the motor system of clarias gariepinus in front of its cerebellum. The stimulation was done by an implanted stainless steel electrode ($100\ \mu m$ diameter, and $1.0\ mm$ depth). The range of stimulation signal was between 2.5KHz to 1MHz, and the base signal (on and off switcher) is between 100Hz to 200Hz, with the duty varied from 5% to 20%. As expected, due to the capacitive properties of the ionic environment at the high frequency, the DC current, which passes through tissue, was almost zero (with 2% precision). In this experiment, the storage charge in each cycle at 200KHz was equal to $3nC$.

   Table 1 shows the minimum discharge in each stimulation when a significant macroscopic response was observed.

*Table 1: The minimum discharge value in each fish brain stimulation, when a meaningful macroscopic response was observed.*

| Type of stimulation | On & off Mean frequency (Hz) | Frequency of stimulation | Duty (%) | Charge amplitude (nc) |
|---|---|---|---|---|
| Random | 200 | 1 MHz | 5 | 0.3 |
| Random | 200 | 200KHz | 5 | 3 |
| Regulate | 200 | 3.5KHz | 5 | 30 |
| Regulate | 150 | 150Hz | 5 | 60 |

As expected, we didn't see any macroscopic response in the direct (continues) ultra-high frequency stimulation. However, the fish motor system was stimulated and the fish swam forward when the stimulation switched on and off at the low frequency (100Hz-200Hz)

2. Weak stimulation of afferent nerves.

    The electrode is the same as in the previous experiment, but its depth was 2mm. Table 2 shows the amount of discharge in each period when a significant macroscopic response was observed. As was mentioned earlier in our model, we expect that at high frequencies the system rapidly adapts itself as was observed in our experiment [we observe that depending on the intensity of stimulation, the subject does not respond absolutely or after a few rotations, it stops the responses]. When the signal frequency is superimposed on a low-frequency carrier wave, we observed a meaningful slowdown in the adaptation. Furthermore, as the model predicts, by randomizing the carrier wave (with a Poisson distribution, with mean 200Hz and 15% duty), the stimulation efficiency was greatly improved, and the motor response increased sharply. Interestingly, a low field intensity stimulation which was carried by a low constant frequency had, sometimes, no effect on the subject, and changing the constant frequency with a random one, caused a smooth response from the subject. movie 1 shows the respond of fish under later stimulation method.

*Table 2: The amount of discharge in each period, when a significant macroscopic response was observed in the fish locomotion.*

| Type of stimulation | On & off Mean frequency (Hz) | Frequency of stimulation | Duty (%) | Charge amplitude (nc) |
|---|---|---|---|---|
| Random | 200 | 300KHz | 15 | 0.7 |
| random | 200 | 2.5KHz | 15 | 12 |
| regulate | 150 | 150Hz | 15 | 140 |

# Conclusion

In recent years there has been increased attention to the ultra-high frequency electrical stimulation of the brain and nervous system. This type of stimulation, due to its less destructive effects, is suitable for medical and cyborg applications. In this article, we provide a brief description of the mechanism underlying the ultra-high frequency stimulation. Then, from an informatics perspective, we proposed a new signaling method for the low amplitude ultra-high frequency stimulation. In this method, we have tried to reduce adaptation of nerve system by modulating the stimulation signal via a low frequency rectangular random wave. By this method, we could reach the "almost zero discharge" with minimum destructive effect. Hence this method could be applicable to the human DBS and other medical electrical stimulation of the nervous system. As we mentioned, a weak intensity stimulation causes an increased probability of neuron firing at the natural environmental noises. We speculate that this phenomenon result in an asymmetric firing rate in the stimulated area compare to other regions however this description needs more investigations.

## Author Contributions

M. J. and Y. J. designed the research, conceived the experiments, and developed the method and theory. The theory and data were analyzed and controlled by M. G. The results were discussed and interpreted by M. J. The manuscript was written and revised by M. G. M. J. and Y. J.

## Additional Information

The authors declare that they have no competing interests.